\documentclass{aa} 
 
\usepackage{xcolor}
\usepackage{graphicx}
\usepackage{txfonts}
\usepackage{lipsum}
\usepackage{subcaption}         
\usepackage{lscape}             
\usepackage{placeins}           

\newcommand{\cmc}{$\rm cm^{-3}$}

\newcommand{\ebv}{E(B-V)}
\newcommand{\starburst}{\texttt{STARBURST99}}
\newcommand{\mmm}{\texttt{M3D}}
\newcommand{\mmmm}{\texttt{M$^3$}}

\newcommand{\mapv}{\texttt{MAPPINGS V}}
\newcommand{\chianti}{\texttt{CHIANTI V.8}}
\newcommand{\map}{\texttt{MAPPINGS}}
\newcommand{\athena}{\texttt{ATHENA++}}
\newcommand{\moc}{\texttt{MOCASSIN$^{\rm 3D}$}}

\newcommand{\ha}{{H$\alpha$}}
\newcommand{\hb}{{H$\beta$}}

\newcommand{\sii}{{[S\sc{ii}]}}

\newcommand{\oi}{{[O\sc{i}]}}
\newcommand{\oiii}{{[O\sc{iii}]}}

\newcommand{\hii}{H{\sc ii}}

\usepackage{natbib}
\usepackage{amsmath}

\begin{document}

\title{The Internal Nebular Attenuation Curve of Three-Dimensional Turbulent HII regions}


%

   \author{Yifei Jin \inst{1}\corrauth{yfjsci@gmail.com}
        \and Yong Shi \inst{1}
        \and Ralph Sutherland\inst{2}
        \and Ziyu Huang\inst{3}
        \and Chuanfei Dong \inst{3,4}
        }

   \institute{Westlake University, 600 Dunyu Road, Xihu District, Hangzhou, Zhejiang 310030 PR China\
            \and Mount Stromlo Observatory,  Australian National University, Weston Creek, Canberra, ACT 2611 Australia\
            \and Department of Astronomy, Boston University, Boston, MA 02215 USA \
            \and School of Natural Sciences, Institute for Advanced Study, Princeton, NJ 08540, USA}

   \date{Received September 30, 20XX}

 
  \abstract{
The internal dust attenuation of the \hii\ region reduces the observed emission-line fluxes.
Turbulent density fields within each \hii\ region change the degree of the line-of-the-sight obscuration of the emission-line fluxes.
In this paper, we implement the dust Monte-Carlo radiative transfer in the latest \mmm\ code, creating the emission-line maps attenuated by the internal turbulent dust obscuration with the varying Mach numbers.
The internal density and temperature fluctuations of \hii\ regions make the radiative transfer of hydrogen lines neither Case A nor Case B conditions, resulting in the global \ha\ to \hb\ ratio of approximately 3.02-3.03, differing from the widely-used value of 2.86.
This deviation from Case B is because the temperature of these \hii\ regions is cooler than 10,000~K.
We further derive the internal nebular attenuation curve from the attenuated Hydrogen lines, finding that the clumpy structures within \hii\ regions do not change the slope of the internal attenuation curve. 
This is because the heavy dust obscuration of dense clumps is canceled out by the high in-situ production of emission-line intensities.
}

                
\maketitle

\nolinenumbers

\section{Introduction}

Ionized nebulae (\hii\ regions)  surrounding young, hot, massive stars are the prominent sites producing the galactic emission-lines.
Analyzing the \hii\ region emission-line properties decodes fundamental properties of the interstellar medium and is a key part of understanding the physical parameters of galaxy formation and evolution scenarios. 
These properties include the gas-phase metallicity \citep{Sanders2021}, the interstellar medium (ISM) pressure/density \citep{Kaasinen2017} and the spectrum of the ionization field \citep{Veilleux1987}.
The correction for the loss of fluxes caused by the internal dust obscuration is critical to these inferences because the nebular regions are preferentially located in dusty star-forming clouds.

An attenuation curve is adopted to parameterize the wavelength-dependent effect of dust obscuration, in which its shape encodes the complexity of dust properties and the spatially-geometrical relation between the stars and their surrounding dust/gas absorbers \citep{Seon2016,Reddy2020}.
The nonuniform distribution of dust around stars varies the column density and the covering fraction of the dust obscuration to the stellar continuum and the nebular emission \citep{Caplan1986}.

In an individual nebula, the total nebular spectrum could be dominated by less dust affected, bright \oiii$\lambda\lambda$4959,5007 and \hb-emitting regions that fill most of the nebular volume, like what is observed in LMC-N59 \citep{Pellegrini2012}. 
In some cases, the nebular spectrum could be dominated by the reddened line species residing at the edge of nebula, like \sii$\lambda\lambda$6717,31 and \oi$\lambda$6300. 
One case is the nebula M16 \citep{Hester1996}, which may be closely associated with the surfaces of dusty dense clumps \citep{Mellema2006}.
In some extreme cases, like DEM 301 \citep{Oey1997}, the nebular attenuation is absent in the regions where the lines-of-sight allow ionizing photons to escape from the nebula without significant absorption \citep{Pellegrini2012}.
In the other extreme case, like ultracompact \hii\ regions \citep{Wood1989}, the nebular emission-lines are heavily embedded in compact dusty clumps where the ionization front is completely neutral \citep{Petrosian1972}.
Therefore, the spectrum of a clumpy \hii\ region is the aggregate of many lines of sight through varying attenuation.
The integrated light at different wavelengths may be biased to the brighter regions at different optical depths.
This is an essential difference between the attenuation of the clumpy and uniform ISM.

The influence of the turbulence in \hii\ regions on the nebular attenuation curve may be at least twofold.
On one hand, the dust attenuation depth in the patchy ISM varies in different directions. 
The dust obscuration is heavier in the directions of the larger column densities, usually with denser clumps along the line of the sight \citep{Zamora-Aviles2019}.
On the other hand, the density fluctuation of the ionized gas changes the intrinsic emission-line fluxes by deviating the temperature and ionization structures from the uniform density models within the \hii\ regions.  
This nonuniform ionization structure is a seldom-considered factor which cannot be reproduced in a symmetric homogeneous model \citep{Jin2022b}.
Substructure is ubiquitous in \hii\ regions as they evolve in their turbulent ambient ISM, so inhomogeneity is inevitable \citep{Elmegreen2004}.

Observations and simulations have revealed that the importance of the turbulent motion in star-forming galaxies is increasing along the redshift \citep{Wisnioski2015,Pillepich2019}.
Especially at high-redshift $z>2$, the high incidents of galaxy mergers further perturb the turbulence in the ISM \citep{Sparre2022}.
The unprecedented Infrared (IR) spectroscopic capability of JWST pushes the frontier of the rest-frame emission-line studies to the very early Universe around $z\sim12$, opening a new window to understand the interplay between the ISM turbulence and the emission-line behaviors as well as the dust attenuation \citep{Shapley2025,Reddy2025}.

To begin to test this qualitative description, in this paper, we create a series of turbulent \hii\ regions with varying Mach numbers, and investigate the impact of the turbulent density fluctuation of the ionized gas on the measurement of the nebular attenuation curve.
We describe turbulent \hii\ region models in Section~\ref{sec:model} and their dust-attenuated hydrogenic-line maps in Section~\ref{sec:Hmap}.
The deduced nebular attenuation curve is reported in Section~\ref{sec:atten_curve}. 
The conclusion is summarized in Section~\ref{sec:conclusion}.
The purpose of this paper is to demonstrate the impact of the non-uniform turbulent density field on the internal attenuation curve of \hii\ regions.

\section{Model}\label{sec:model}

We create a series of \hii\ regions with the turbulent density fields of varying Mach numbers through the following steps.
We first run hydrodynamic simulations to create the turbulent boxes with varying Mach numbers.
We then illuminate the turbulent boxes using the Monte-Carlo radiative transfer photoionization code to create cubes of nebular emission-lines.
The emission-lines are reddened by the dust attenuation that is proportional to the intrinsic density fluctuations of the turbulent box along the line of sight.

\subsection{Turbulent Interstellar Medium}

We use the \athena\ code \citep{Stone2020} to create the turbulent boxes, which provide the density fields for post-processing.
The density fields are characterized by the Mach number  and the lowest wavenumber of the turbulent box $k_{min}$, respectively.
The Mach number controls the dispersion of the probability distribution function (PDF) of the gaseous density \citep{Monlia2012}. 
The $k$ parameter defines the maximal periods of the density oscillation along each side of the turbulent boxes, which constrains the largest scale of the turbulent filament in the simulation. 
The physical domain size is $L=$ 64 pc and the simulations are performed on a uniform grid with a resolution of $N^3=125 \times 125 \times 125$ cells.
The turbulence boxes with different Mach numbers are the selected snapshots from simulations with identical driving, $\dot{E} / \bar{\rho} L^2 c_s^3=10$, where $\bar{\rho}$ is the average density, $L$ is the size of the domain and $c_s$ is the sound speed.
We create the turbulent boxes of the Mach numbers $\mathcal{M}=$0.997, 2.484, 4.739, and 8.791 with the lowest wavenumbers $k_{min}=$ 4, and $\mathcal{M}=$0.997, 2.472, 4.856, and 8.817 with the lowest wavenumbers $k_{min}=$ 9.
The chaos of the gas increases as the time evolving and the energy cascades to the small scale and dissipates at the small eddies. 
The clumpy and filamentary structures appear during the nonlinear interaction of gas flow.  The mean density of the turbulent medium is 10~\cmc.

\subsection{Photoionization Models}

The \mmm\ code\footnote{https://github.com/Jinyifei/M3} \citep[the updated version of \mmmm , see][]{Jin2022a} is a self-consistent three-dimensional Monte-Carlo radiative transfer (MCRT) photoionization code designed to create \hii\ regions with arbitrary geometries.
It is the descendant of the \map\ project, which was initialized by \cite{Dopita1976} and evolved through the continuous development of \cite{Binette1985,Sutherland1993} and \cite{Groves2004}.
The \mmm\ code inherits the framework  of \mapv , the most up-to-date version of \map\ code that was restructured by \cite{Sutherland2018}, which includes the updated microphysics of the ISM, the efficient strategies of the cooling and heating calculations, and the \chianti\ atomic data \citep{DelZanna2015}.

In the latest version of \mmm, we consider the absorption and scattering by gas and dust by comparing the total random optical depth, $\tau_p$, with the analytical optical depth, $\tau_l$, over each displacement of $l$,
\begin{equation}
    \tau_l = (\kappa_{abs}^{gas}+\kappa_{abs}^{dust}+\kappa_{scat}^{dust})\cdot \rho \cdot l,
\end{equation}
where $\kappa_{abs}^{gas}$, $\kappa_{abs}^{dust}$ and $\kappa_{scat}^{dust}$ are the coefficients of gas and dust absorption and dust scattering, and $\rho$ is the gas density.
Once the total analytical optical depth is larger than the random optical depth, the photon is then determined to be either scattered or absorbed according to the probability of each event.
\begin{align}
  P_{scat}^{dust} = \frac{\kappa_{scat}^{dust}}{\kappa_{tot}} \label{p_scat}  \\
  P_{abs}^{gas} = \frac{\kappa_{abs}^{gas}}{\kappa_{tot}} \label{p_gas_abs}  \\
  P_{abs}^{dust} = \frac{\kappa_{abs}^{dust}}{\kappa_{tot}} \label{p_dust_abs} 
\end{align}
The dust scattering only changes the traveling directions of photon packets.
Once the photon is absorbed by a dust grain, it is re-emitted as IR photons without further impact on the successive radiative transfer process.
This technique is also used in other MCRT codes, like \moc \citep{Ercolano2005}.

We adopt the fully sampled stellar population to obtain the shape of the ionizing spectrum and scale it to match with the desired ionizing photon rate.
The stellar ionizing spectrum is generated by \starburst\ with a continuous star formation history with a constant star formation rate of 1~M$_\sun$~yr$^{-1}$ over a timescale of 5~Myr.
We adopt a Salpeter initial mass function \citep[IMF,][]{Salpeter1955} and the Geneva stellar evolution track \citep{Lejeune2001} with the  Wolf-Rayet atmosphere library of \cite{Schmutz1992}.
The stellar abundance is aligned with the gas-phase metallicity by the same fractional solar abundance \citep{Asplund2009}.
The ionizing source is placed at the center of the domain and the total luminosity of the ionizing source is set to be 10$^{40}$~erg~s$^{-1}$.

\subsection{Dust grains}

The composition of different sized dust grain is the key factor to determine the absorption and scattering contribution of dust.
The grain size distribution is the balance of the processes relevant to the dust destruction and formation.
We adopt the grain distribution proposed by \cite{Mathis1977},
\begin{equation}
    dN(a)/da=ka^{-\alpha},
\end{equation}
where $N(a)$ is the number density of the dust grain at the size $a$, and $\alpha$ is selected to be 3.5.

In the dusty \hii\ regions, the heavy elements are depleted from the gas-phase ISM and incorporated into the dust grains.
The atomic depletion pattern is a factor constraining the dust attenuation \citep{Salim2020}.
We adopt the uniform depletion factor given by \cite{Jenkins2009} for the model grids with varying metallicity, although \cite{Savage1996} has pointed out that the dust depletion is dependent on the ISM environment.
Because the fractional depletion does not change with the metallicity of the gas, the gas-to-dust
mass ratio is in these models is only proportional to the gas-phase metallicity.

\section{Dust Attenuation of Emission-Lines}

We produce the 2D emission-line maps from the 3D data cube by integrating the emission-line fluxes reddened by the dust attenuation along the line-of-the-sight (LOS).
The dust attenuation is constrained by the color excess $\ebv$ and the dust attenuation curve. 
We first derive the color excess \ebv\ cube from the density field by assuming that a constant ratio between the \ebv\ and the hydrogen column density, $N_H$,
\begin{equation}
    E(B-V)/N_H=\frac{4\times10^{-22}~cm^{-2}}{2.040685~mag}, \label{eq:ebv}
\end{equation}
which is adapted from \cite{Zucker2021} and \cite{Draine2009}.
The $N_H=n_H\times dr$ is the column density of the cell with volume number density of $n_H$ and size $dr$.
This is the reverse approach of deriving the 3D density field of the Local Bubble \citep{O'Neill2024} from the 3D dust attenuation map \citep{Edenhofer2024}.

The attenuation $A_\lambda$ sensed by the $i^{th}$ cell is the integration of the attenuation of the $j^{th}$ cells on the line-of-the-sight, which is derived by:
\begin{equation}    
    A_\lambda^{i} = \sum_j E(B-V)_{j} \cdot x(\lambda),
\end{equation}
where $A_\lambda^{i}$ is the integrated attenuation of the $i^{th}$ cell, $E(B-V)_{j}$ is the color excess of the $j^{th}$ cell, and $x(\lambda)$ is the intrinsic extinction curve,  which is the curve proposed by \cite{Fitzpatrick1999}.

The final projected 2D emission-line map is the integration of the dust reddened emission-line fluxes along the LOS:
\begin{equation}
    f_\lambda^{proj} = \sum_{i} f_\lambda^{i} \times 10^{-0.4 \times A_\lambda^{i}},
\end{equation}
where the $f_\lambda^{proj}$ is the projected line flux at the wavelength $\lambda$, and the $f_
\lambda^{i}$ is the intrinsic flux in the $i^{th}$ cell.

Figure~\ref{fig:k4map} and \ref{fig:k9map} present the projected \ebv\ maps of the turbulent models. 
The \ebv\ values scatter within each model because of the density fluctuation of the turbulent ISM.
The fluctuation of the \ebv\ map increases as the increasing Mach number of the turbulence.
The \ebv\ maps of the $k=4$ models are more filamentary than the $k=9$ models.

\section{The Hydrogen Lines}\label{sec:Hmap}

We compute the Hydrogen lines from the Lyman series (energy level $n=1$) to the Further series (energy level $n=6$) based on the hydrogen recombination coefficients given by \cite{Storey1995}.
The transfer of the hydrogen resonance lines is assumed to be a linear combination of the Case A and the Case B conditions, based on the local temperature and density \citep{Sutherland1993,Jin2025}.

\begin{figure*}
    \centering
    \includegraphics[width=\linewidth]{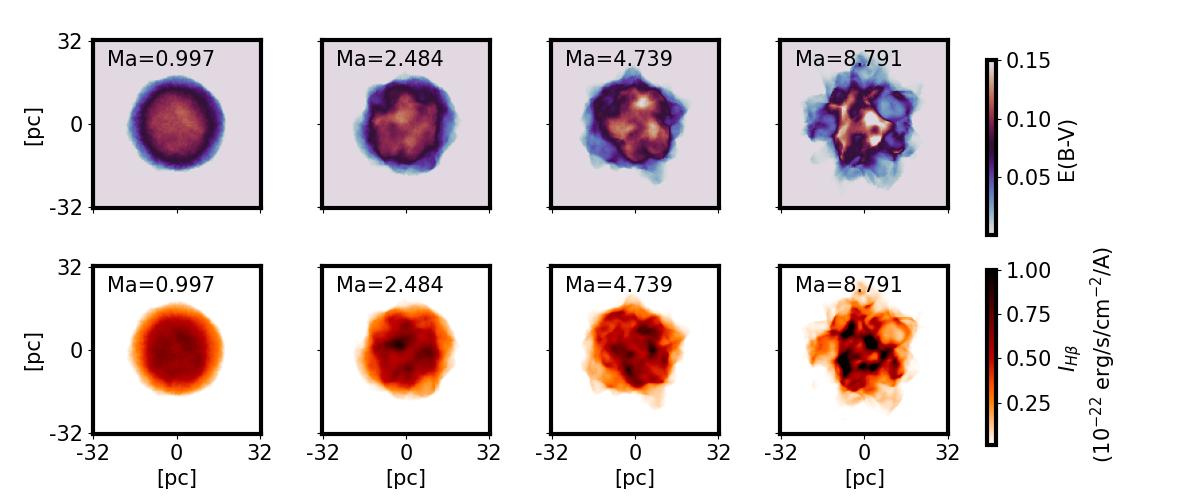}
    \caption{The maps of the projected \ebv\ and the attenuated \hb\ intensity of the $k=4$ and $\mathcal{M}$=0.997, 2.484, 4.739, 8.791 models.}\label{fig:k4map}
\end{figure*}

\begin{figure*}
    \centering
    \includegraphics[width=\linewidth]{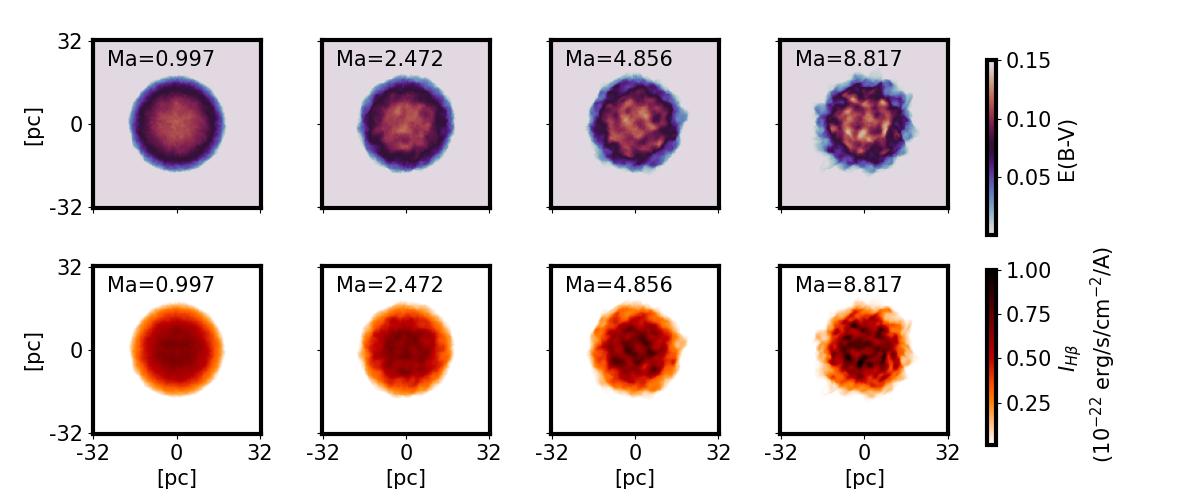}
    \caption{Same to Figure~\ref{fig:k4map} but for the $k=9$ and $\mathcal{M}$=0.997, 2.472, 4.856, and 8.817 models.}\label{fig:k9map}
 \end{figure*}

In Figure~\ref{fig:k4map} and \ref{fig:k9map}, we present the projected \hb\ maps of the turbulent models with varying Mach numbers and $k$ parameters.
In the low Mach number models, the density field of the \hii\ region is smooth so the \hb\ distribution is close to the \hii\ region with the smooth spherical geometry.
In the high Mach number models, the density field of the \hii\ region becomes clumpy as the contrast between the low-density and high-density clumps is getting large.
The \hb\ map of the \hii\ region presents the bright clouds and the \hb\ brightness shows non-symmetric fluctuation.
The boundary of the \hii\ regions is fractal because the ionization front is broken into fragmental by the dense clumps.

The $k$ parameter affects the morphology of \hii\ regions by controlling the filamentary structures in the turbulent box.  We note in passing that it is expected that the heating and expansion flow caused by the \hii\ region evolution may modify the turbulent spectrum of the initial cold starforming gas, although this short experiment should prove indicative of the effects of `clumpiness' compared to previous smooth symmetrical models.
In a fixed-sized box, the larger $k$ parameter breaks the large filaments into smaller pieces, increasing the degree of the clumpiness of the density field.
For the \hii\ regions with the same Mach number, the $k=4$ models show more loops and filaments than the more uniform (on average) $k=9$ models.

\begin{figure*}
    \centering
    \includegraphics[width=\linewidth]{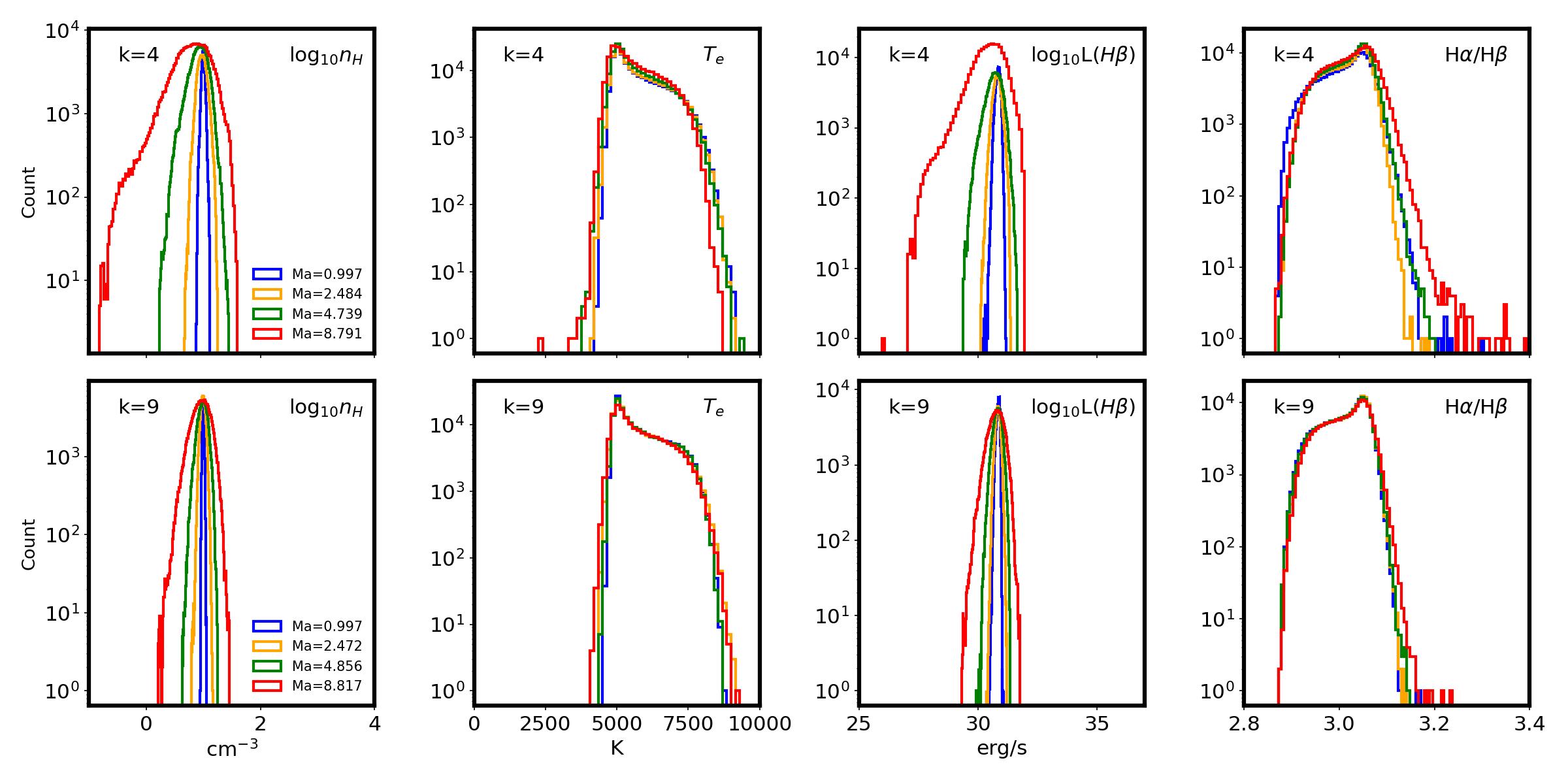}
    \caption{The probability distribution function of the volume number density, $n_H$, the electron temperature, $T_e$, the \hb\ intensity, and the intrinsic \ha/\hb\ ratio of the \hii\ regions of different Mach-number boxes.}\label{fig:pdf}
\end{figure*}

The probability distribution of the emissivity of the hydrogen recombination-lines is affected by Mach numbers through the PDF of the gaseous density.
As shown in Figure~\ref{fig:pdf}, the PDFs of the emissivity of \hb\ are broader in the larger Mach-number turbulent boxes than those in the smaller Mach-number boxes of the same $k$ parameter.
This change is driven by the density variation within each \hii\ region instead of the temperature variation because the PDF of electron temperature barely changes along with the Mach number of the turbulent medium.

\begin{figure*}    
    \centering
    \includegraphics[width=\linewidth]{ 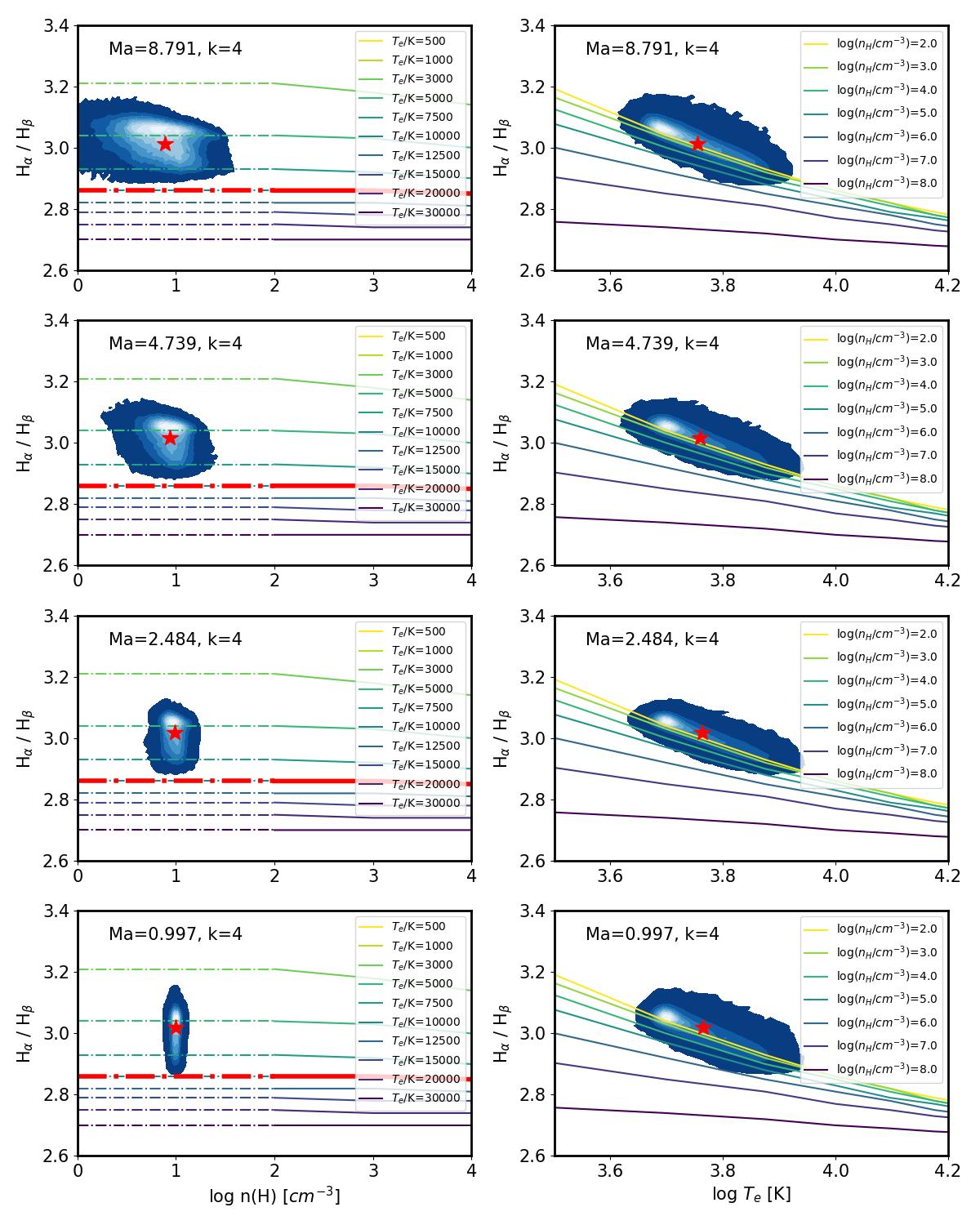}
    \caption{The dependence of the intrinsic \ha/\hb\ ratio on the electron temperature and the gas density. The colored lines are the theoretical ratios in the Case B condition from the Appendix in \cite{Dopita2003}. The red thick line indicates the typical Case B \ha/\hb\ ratios at $T_e=10,000~K$. The dashed line in the left panel are the exploration to the gas with density $n_H < 100$~\cmc. The color-filled contours are the \ha/\hb\ ratio of the turbulent box with Mach number $\mathcal{M}=$0.997, 2.484, 4.739 and 8.791. The $k$ value is 4. The red star indicates the ratio of the integrated \ha\ and \hb\ flux versus the average temperature and density.}
    \label{fig:hahb_k4}
\end{figure*}

\begin{figure*}    
    \centering
    \includegraphics[width=\linewidth]{ 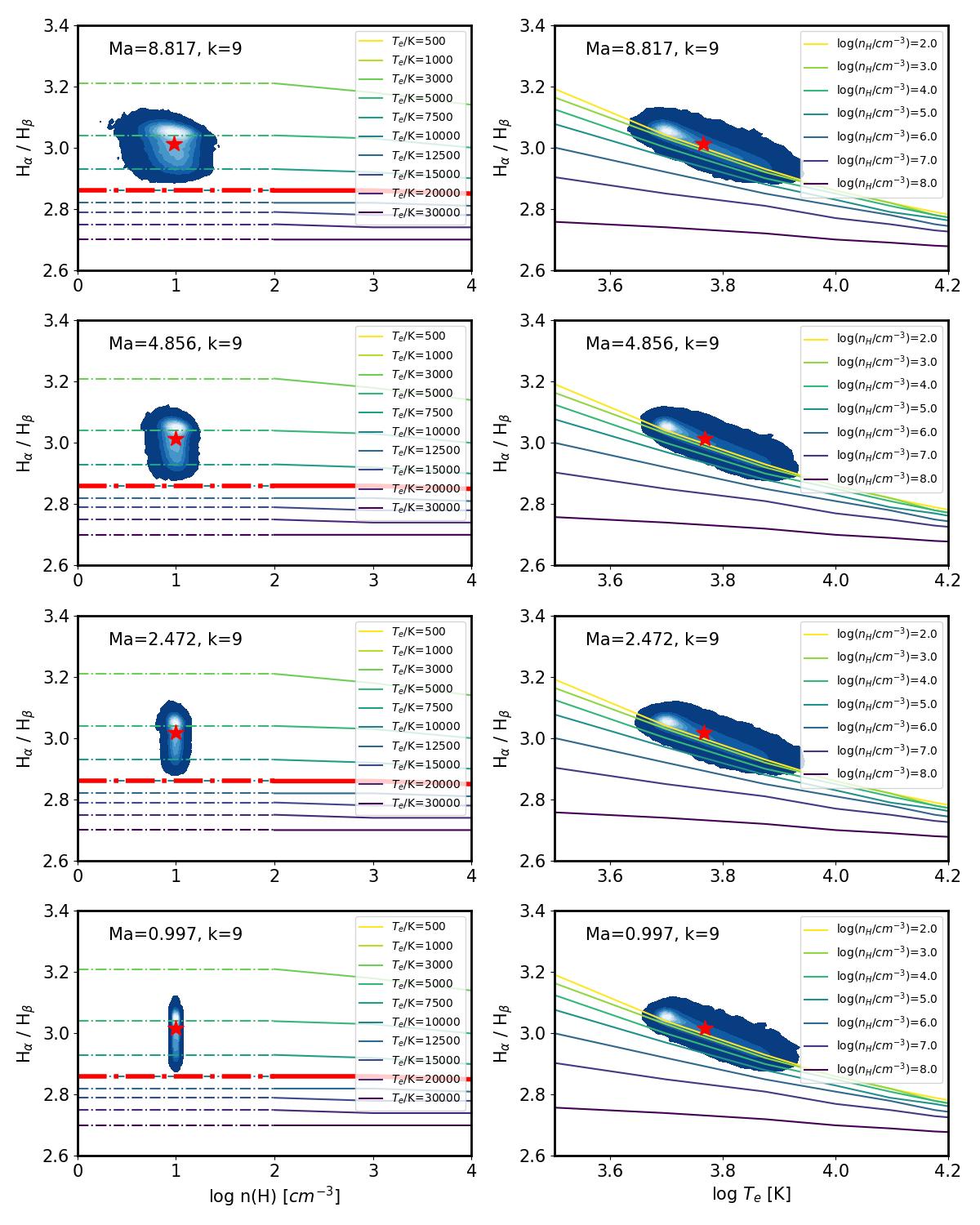}
    \caption{The same figures as Figure~\ref{fig:hahb_k4} but for the $k=9$ and $\mathcal{M}$=0.997, 2.472, 4.856, and 8.817 turbulent boxes.}
    \label{fig:hahb_k9}
\end{figure*}

The intrinsic \ha/\hb\ ratios vary within \hii\ regions.
Figure~\ref{fig:pdf} also shows the probability distribution of the intrinsic \ha/\hb\ ratios within each \hii\ region.
The medium values of \ha/\hb\ ratio in the turbulent \hii\ regions are between 3.02 and 3.03, deviated from the standard ``Case B'' value of 2.86. 
The PDFs of the \ha/\hb\ ratio have the similar shape between different Mach numbers.
Within each \hii\ region, 68.27 per cent of pixels have \ha/\hb\ ratios falling between 2.96 and 3.06.
There is a subtle difference in the 3$\sigma$ variation of PDF of the \ha/\hb\ ratio between the boxes of the same $k$ parameter with different Mach numbers.
The 3$\sigma$ variation is enhanced in turbulence with the larger Mach numbers than the smaller Mach numbers for both $k=4$ and 9 turbulence.
As shown in Figure~\ref{fig:hahb_k4} and \ref{fig:hahb_k9}, these variations of \ha/\hb\ ratio are driven by electron temperature but independent of density, as seen in both the data and theoretical grids.

\section{The Nebular Attenuation Curve}\label{sec:atten_curve}

\begin{figure*}
    \centering
    \includegraphics[width=\linewidth]{ 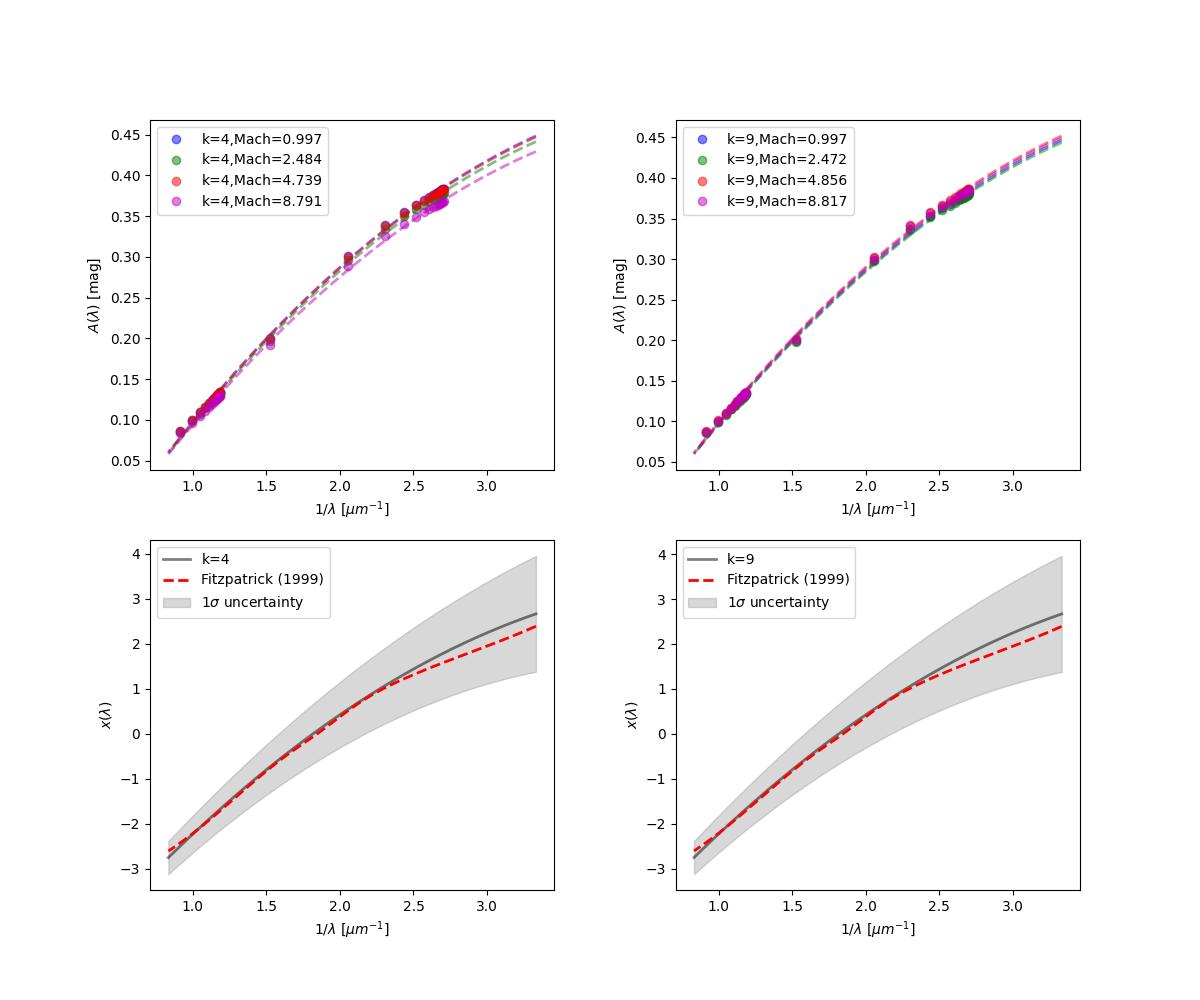}
    \caption{\textbf{Upper:} The attenuation as a function of $1/\lambda$ for the $k=4$ (left) and $9$ (right) turbulent boxes. The dots are the attenuation values derived from individual Hydrogen lines by Equation~\ref{eq:atten} and the lines are the best-fit attenuation functions of different Mach number turbulent boxes. \textbf{Lower:} The derived attenuation curves for the $k=4$ (left) and $9$ (right) turbulent boxes. The black lines are the best-fit curves and the gray shades are the 1-$\sigma$ uncertainty areas of the fitting. The red dashed lines is the attenuation curve of the \cite{Fitzpatrick1999}.}
    \label{fig:kcurve}
\end{figure*}

We derive the attenuation of the Hydrogen lines by comparing the attenuated line fluxes to their intrinsic fluxes.
\begin{equation*}
    A_\lambda^{H} = -2.5 \log_{10} \frac{F_\lambda^{redden}}{F_\lambda^{intrinsic}},
\end{equation*}\label{eq:atten}
where the $F_\lambda$ is the flux of different Hydrogen lines, and the $A_\lambda^{H}$ is the discrete attenuation at each hydrogen-line wavelength.
Given we only consider the Hydrogen lines from the Lyman series to the Further series with 15 lines of each series, there are 90 sampling of the attenuation values as a function of the wavelength.
We use the two-degree polynomial function to fit the full wavelength-dependent attenuation function.
\begin{equation*}
    A_\lambda = c_0 + c_1 \lambda^{-1} + c_2 \lambda^{-2},
\end{equation*}\label{eq:attenfit}
The final nebular attenuation curve is formulated as below:
\begin{equation*}
    x(\lambda) = \frac{A_\lambda}{A_B - A_V} = \frac{A_\lambda}{E(B-V)},
\end{equation*}\label{eq:kcurve}
where the $A_\lambda$ is the fitted attenuation at the wavelength $\lambda$.
We equate the $A_V$ with the attenuation at 5500\AA\ and the $A_B$ with the attenuation at 4400\AA . 
The fitted attenuation curves are shown in Figure~\ref{fig:kcurve} and the best-fit coefficients are listed in Table~\ref{tab:kcurve}.

As shown in Figure~\ref{fig:kcurve}, the $A_\lambda$ values vary with the Mach number largely at the wavelength range shorter than $\lambda\lesssim6600$\AA\ ($1/\lambda\gtrsim1.5\mu m^{-1}$) but remain similar at the long-wavelength range.
This variation is caused by the change of the intrinsic \ebv\ value of the ionized gas, which is more pronounced in the $k=4$ turbulent boxes than in the $k=9$ turbulent boxes, as shown in Table~\ref{tab:ebv_comparison}.
After removing the effect of the varying \ebv, the derived attenuation curves of all the models, $x(\lambda)$, are consistent with the essential \cite{Fitzpatrick1999} curve within the 1-$\sigma$ uncertainty.


\begin{table*}[htbp]
\centering
\begin{tabular}{c 
                r@{ $\pm$ }l 
                r@{ $\pm$ }l 
                r@{ $\pm$ }l 
                r@{ $\pm$ }l}
\hline
\rule[-3mm]{0mm}{8mm}
Mach number
 & \multicolumn{2}{c}{$c_{0}$} 
 & \multicolumn{2}{c}{$c_{1}$} 
 & \multicolumn{2}{c}{$c_{2}$} 
 & \multicolumn{2}{c}{$R_V$} \\
\hline
\multicolumn{9}{c}{$k=4$} \\
\hline
0.997 & -5.702 & 0.185 & 3.884 & 0.183 & -0.412 & 0.045 & 3.589 & 0.051 \\
2.484 & -5.707 & 0.185 & 3.890 & 0.183 & -0.413 & 0.045 & 3.593 & 0.051 \\
4.739 & -5.710 & 0.185 & 3.893 & 0.183 & -0.414 & 0.045 & 3.594 & 0.051 \\
8.791 & -5.720 & 0.184 & 3.903 & 0.183 & -0.416 & 0.045 & 3.604 & 0.051 \\
\hline
\multicolumn{9}{c}{$k=9$} \\
\hline
0.997 & -5.695 & 0.185 & 3.878 & 0.184 & -0.410 & 0.045 & 3.584 & 0.051 \\
2.472 & -5.697 & 0.185 & 3.880 & 0.183 & -0.411 & 0.045 & 3.586 & 0.051 \\
4.856 & -5.706 & 0.185 & 3.889 & 0.183 & -0.413 & 0.045 & 3.592 & 0.051 \\
8.817 & -5.711 & 0.185 & 3.894 & 0.183 & -0.414 & 0.045 & 3.596 & 0.051 \\
\hline
\end{tabular}
\caption{Best-fit coefficients $c_{0}, c_{1}, c_{2}$ and derived $\rm R_V$ with $1\sigma$ errors for different Mach number turbulent boxes of $k=4$ and $9$.}
\label{tab:kcurve}
\end{table*}

\begin{table*}[htbp]
\centering
\begin{tabular}{c cc}
\hline
\rule[-2mm]{0mm}{8mm}
Mach number & $E(B-V)_{\text{fit}}$ & $E(B-V)_{\text{mean}}$  \\
\hline
 \multicolumn{3}{c}{\rule[-2mm]{0mm}{5mm}$k=4$} \\
\hline
0.997 & $0.072 \pm 0.001$ & 0.076 \\
2.484 & $0.071 \pm 0.001$ & 0.074 \\
4.739 & $0.071 \pm 0.001$ & 0.070 \\
8.791 & $0.069 \pm 0.001$ & 0.051 \\
\hline
 \multicolumn{3}{c}{\rule[-2mm]{0mm}{5mm}$k=9$}  \\
\hline
0.997 & $0.071 \pm 0.001$ & 0.077 \\
2.472 & $0.071 \pm 0.001$ & 0.077 \\
4.856 & $0.072 \pm 0.001$ & 0.075 \\
8.817 & $0.072 \pm 0.001$ & 0.071 \\

\hline
\end{tabular}
\caption{Comparison of $E(B-V)_{\text{fit}}$ (with $1\sigma$ uncertainties) and mean $E(B-V)$ for $k=9$ and $k=4$ across Mach numbers. The values are extracted from the ionized gas only in photoionized models.}
\label{tab:ebv_comparison}
\end{table*}

\section{Discussion}
In the photoionization models, the dust-to-gas ratio is always assumed to be a fixed value across the entire photoionized region \citep{Dopita2000,Groves2004,Dopita2006a}. 
We follow this assumption and select the dust-to-gas ratio in the ionized gas the same to the ratio in the neutral molecular clouds.
Under this assumption, both the color excess, \ebv , and the emission-line intensity are proportional to the density of the total hydrogen atom.
Although the column density and the dust covering factor vary with different Mach numbers in turbulent density fields, the heavy dust obscuration in one direction is canceled out by the high emissivity produced by the same groups of dense clouds.
Therefore, the derived global attenuation curve has no pronounced dependence on the Mach number of the nebular internal turbulence.  

Although the global attenuation is identical across boxes with different Mach numbers, the spatially-resolved projected attenuation maps reveal the density structures of the turbulence.
The attenuation maps appear smooth in low-Mach-number boxes, where the density fluctuations are small, while more clumpy structures are shown in the high-Mach-number boxes.
The attenuation maps also trace the shells and filaments present in the density fields.
Particularly, the $\mathcal{M}=8.791$, $k=4$ attenuation map exhibits the bubble-like structures which arise from the interplay between the turbulent driving scale and the Mach number.
These bubbles are seen in external galaxies \citep{Watkins2023} and in our Milky-way, especially the Local Bubble \citep{Zucker2021}, which has been expanding over 14~Myrs \citep{O'Neill2024}.

In our models, we use the turbulent boxes without the contribution of gravity to generate the density fields.
In galactic hydrodynamic simulations, gravity is a fundamental parameter to concentrate gas extending the high-density tail in the density probability function \citep{Burkhart2019}. 
This changes of density fluctuation may increase the 2D variations on attenuation maps but have subtle changes in the measurement of the global attenuation curve.

In our models, we ignore the dust drifting effect caused by the radiation pressure acting on the dust grains.
Under the static equilibrium condition, the radiation pressure causes a radial gradient of the density within the ionized gas \citep{Dopita2003b,Dopita2006}, compressing the dust and gas into an ionized shell, resulting in an elevated dust-to-gas ratio in the outer region of \hii\ regions \citep{Draine2011}.
By including the dust drifting effect, the major fraction of the line emissivity is expected from the inner part of \hii\ regions, separating from the attenuation of the dust whose large fraction resides at the outer part of \hii\ regions.
The action of the radiation pressure on dust will be discussed in the future version of \mmm.

Apart from the internal dust attenuation of nebula, the observed flux suffers from the dust attenuation in-between the \hii\ regions \citep{Salim2020}.
These inter-\hii\ region space contributes the global dust attenuation without the high nebular emissivity.
The radiative transfer models of dusty turbulent ISM have shown that the dusty clumps in the inter-\hii\ region are one driver of the variation of the global attenuation curve in galaxies \citep{Witt2000,Seon2016}.
The relative position of the \hii\ regions and the dust in-between (so-called dust-star geometry) alters the slope of dust attenuation curve \citep{Wild2011,Narayanan2018,Reddy2025}.
Our results further imply that the observed variation of the galactic-scale attenuation curve is caused by the dusty turbulent gas in-between the \hii\ regions rather than the gas within the \hii\ regions.


\section{Conclusion}\label{sec:conclusion}

We implement the dust radiative transfer in the latest \mmm\ code and produce the three-dimensional \hii\ regions in the turbulent ISM with varying Mach numbers and wavenumbers.
We derive the attenuation curve from the different series of Hydrogen lines of each \hii\ region.
We find the following conclusions.

\begin{itemize}
    \item The intrinsic Hydrogen line ratios vary within the individual \hii\ region because of the internal density fluctuation caused by turbulence.
    The global \ha/\hb\ ratio is around 3.02 to 3.03, deviating from the ``Case B'' assumption of 2.86. 
    This value does not change with the varying Mach number turbulent structures.
    \item The clumpy and filament structures increase the attenuation of the \hii\ region. However, the derived integrated selective attenuation curve is the same to the intrinsic input \cite{Fitzpatrick1999} curve.
    The density fluctuation has no impact on the integrated attenuation curve of the turbulent \hii\ region. 
    \item The best-fit \ebv\ is consistent with the mean \ebv\ of the ionized gas. However, the mean \ebv\ varies with Mach numbers and wavenumbers.
\end{itemize}

Our results highlight the consistency of the attenuation curves derived from the observed integrated emission--lines and the intrinsic curves, which is important for the future discussions of the dust attenuation of the nebulae in turbulent environments observed by the SDSS-V/LVM survey.

\begin{acknowledgements}

Y.F.J. acknowledges the computational resources at the Westlake High-Performance Computing Center.
Y.F.J. and Y.S. acknowledge the support from the Startup funding of New-joined PI of Westlake University.
This work is supported by the National
Natural Science Foundation of China (NSFC grants 12141301, 12121003, 12333002).
C.D. and Z.H. acknowledge support from DOE grant DE-SC0024639, the Alfred P. Sloan Research Fellowship, and the IBM Einstein Fellow Fund at the Institute for Advanced Study, Princeton.
C.D. and Z.H. would like to acknowledge the high-performance computing support from National Energy Research Scientific Computing Center, a DOE Office of Science user facility.

\end{acknowledgements}

%
 
 \bibliographystyle{aa} 
 \bibliography{reference}{} 


\end{document}